\documentclass[%
 reprint,
 amsmath,amssymb,
 aps,
]{revtex4-2}

\usepackage{graphicx}
\usepackage{dcolumn}
\usepackage{bm}
\usepackage[linktocpage,colorlinks=true,linkcolor=blue,citecolor=blue,urlcolor=blue,breaklinks=true]{hyperref}   

\def\u{{\mathbf{u}}}
\def\ldef{\mathrel{\mathop:}=}

\def\mysec#1{{\bf #1 --}}

\begin{document}

\title{A New Circle Theorem for Two Dimensional Ising Spin Glasses}

\author{Chaoming Song}%
\email{c.song@miami.edu}
\affiliation{%
Department of Physics, University of Miami, Coral Gables FL, 33146 USA.
}%

\date{\today}

\begin{abstract}
The Lee-Yang circle theorem revolutionized our understanding of phase transitions in ferromagnetic systems by showing that the complex zeros of partition functions lie on the unit circle, with criticality arising as these zeros approach the real axis in the thermodynamic limit. However, in frustrated systems such as antiferromagnets and spin glasses, the zeros deviate from this structure, making it challenging to extend the Lee-Yang theory to disordered systems. In this work, we establish a new circle theorem for two-dimensional Ising spin glasses, proving that the square of the partition function exhibits zeros densely packed along the unit circle. Numerical simulations on the square lattice confirm our theoretical predictions, demonstrating the validity of the circle law for quenched disorder. Furthermore, our results uncover a finite-temperature crossover in $\pm J$ spin glasses, characterized by the emergence of a spectral gap in the angular distribution of zeros. This result extends the Lee-Yang framework to disordered systems, offering new insights into spin-glass criticality.
\end{abstract}

\maketitle

\mysec{Introduction}
The study of phase transitions in statistical mechanics has been profoundly influenced by the Lee-Yang circle theorem, which provides fundamental insights into the location of partition function zeros in complex plane analyses \cite{yang1952statistical,lee1952statistical}. In particular, it states that for ferromagnetic Ising models, the partition function satisfies  
\begin{equation}\label{eq:leeyang}
    Z \sim \prod_i (z - z_i),
\end{equation}
where \( z \equiv e^{-2\beta h} \) and the complex roots \( z_i \) lie on the unit circle \( |z_i| = 1 \). This elegant result underpins our understanding of how, in the thermodynamic limit, these zeros approach the real axis and drive criticality, providing a universal understanding of phase transitions.

On the other hand, spin glasses exhibit complex energy landscapes due to quenched disorder, leading to unique phase transition behaviors that remain poorly understood, particularly in low-dimensional systems~\cite{mezard1987spin}. Unlike conventional ferromagnets, where the Lee-Yang zeros form a well-structured circular distribution, spin-glass systems exhibit a far more intricate arrangement of zeros. This complexity arises because the Lee–Yang theorem relies on the positivity of interactions, a property that breaks down in disordered or frustrated systems such as spin glasses, where random or negative couplings spoil the required monotonicity. Instead, numerical simulations have revealed complex patterns for partition function zeros \cite{ozeki1988distribution,matsuda2008distribution}. Despite considerable numerical and theoretical efforts, no analogous {\it circle law} has been rigorously established for frustrated systems, leaving major gaps in our understanding of their phase transitions and critical behavior.

In this work, we introduce a new circle theorem tailored to two-dimensional Ising spin glasses in the absence of an external field (\( z = 1 \)). Contrary to the standard case where \(Z\) factorizes as in Eq.~\eqref{eq:leeyang}, we show that
\begin{equation}\label{eq:circle}
    Z^2 \sim \prod_i (1 - z_i),
\end{equation}
and crucially, all \( |z_i| = 1 \) under broad conditions of quenched disorder on planar graphs. Our exact derivation reveals a hidden geometric constraint on the partition function zeros that persists despite frustration. Similar to Lee–Yang theory, our ``cousin" circle law suggests that the phase transition occurs when the zeros pinch the real axis. We verify our theory numerically in square-lattice spin glasses.

Additionally, our numerical results suggest that in the case of \(\pm J\) disorder, another phase transition emerges at a finite temperature, evidenced by the formation of a {\it spectral gap} in the angular distribution of zeros. This crossover phase may represent a precursor to the genuine glass state at zero temperature. Beyond providing a rare exact analytical handle on spin-glass criticality in finite dimensions, our findings strengthen the conceptual link between complex-plane analyses of partition functions and emergent order in disordered materials. 

\mysec{Self-dual formula}
We consider the Ising spin glass defined on a planar graph $G = (V, E)$ with $n = |V|$ and $m = |E|$, with the Hamiltonian
\begin{equation}
    H = -\sum_{(i,j)\in E} J_{ij} \sigma_i\sigma_j,  
    \notag
\end{equation}
where the coupling constant $J_e$ is assigned to each edge $e \in E$. We reparameterize the couplings in terms of a set of weights $\u = \{ u_e \mid e\in E\}$, where $u_e \ldef \tanh(\beta J_e)$.

Our approach builds on our recent discovery of a new combinatorial self-dual formula for the partition function \cite{song2024kramers}. Below, we briefly review this formula (see Appendix Section~\ref{sec:self} for details). The combinatorial approach of the partition function dates back to the work of Kac and Ward \cite{kac1952combinatorial}, who provided an alternative derivation of Onsager's free energy. Their method is formulated in terms of path integrals
\cite{sherman1960combinatorial,burgoyne1963remarks,sherman1962combinatorial,sherman1963addendum}, 
\begin{equation}
    Z^2 = 2^{2n} \prod_e(1-u_e^2)^{-1} \zeta_F(G,\u)^{-1},
\label{eq:path}
\end{equation}
where $\zeta_F(G,\u)^{-1}$ is defined as an Euler product over all oriented prime cycles,
\begin{align}\label{eq:Fzeta}
\zeta_F(G,\u)^{-1} \equiv \prod_{[p]} \left( 1- (-1)^{w(p)}\prod_{e\in p} u_e \right),
\end{align}
with $w(p)$ representing the winding number of the prime cycle $p$. This identity draws an analogy with the Riemann zeta function. The subscript $F$ highlights the fermionic nature of the Ising model, which assigns a negative weight to cycles with an odd winding number. Based on Eq.~(\ref{eq:Fzeta}), Kac and Ward demonstrated that
\begin{equation}\label{eq:Fedge}
\zeta_F(G,\u)^{-1} = \det \left(I_{2m} - T_{KW}(\u)\right),
\end{equation}
where the Kac-Ward (KW) operator $T_{KW}$ is a $2m\times 2m$ matrix, defined as
\begin{align}
    (T_{KW})_{e',e} = u_e e^{i\alpha(e',e)/2} \notag
\end{align}
if edge $e'$ follows $e$ without backtracking, where $\alpha(e',e)$ is the exterior angle from $e$ to $e'$. This condition ensures that the summation of half exterior angles results in a total phase change of $\pi$ over a cycle, effectively capturing the fermionic sign in Eq.~(\ref{eq:Fzeta}).

\begin{figure}
  \includegraphics[width=1\linewidth]{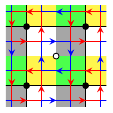}
  \caption{(a) The embedding of both $G$ and its dual $G^*$. The curl operators $D(G,\u)$ (red) and $D(G^*,\u^*)^\dag$ (blue) act as curl operators around the vertex $v$ (black) and its dual $v^{*}$ (white), respectively.}
  \label{fig:square}
\end{figure}

Instead of using the Kac-Ward determinant, we discovered a new determinant formula that explicitly manifests the Kramers–Wannier (KW) duality~\cite{kramers1941statistics}. To achieve this, we embed the dual graph $G^*$ over $G$. In this arrangement, each vertex of $G$ is located inside a face of $G^*$ and vice versa; each edge in $G$ intersects with the corresponding edge in $G^*$. For technical convenience, we require these intersections to be perpendicular. Figure~\ref{fig:square}a illustrates such an embedding for the square lattice, where black and white nodes represent $G$ and $G^*$, respectively. 

A crucial element in our formulation involves the quadrilaterals (colored domains in Fig.~\ref{fig:square}). Each quadrilateral is formed by two neighboring edges in both $G$ and $G^*$, along with a vertex pair $(v,v^*)$, where $v \in V$ and $v^* \in V^*$. The collection of $2m$ quadrilaterals, which collectively tile the entire plane, plays a critical role in our new formulation. Using this setup, we have shown (also see Appendix Section \ref{sec:self}) that
\begin{align}\label{eq:mine1}
    \zeta_F^{-1}(G,\u) = 2^{-n} \prod_{e\in E}(1+u_e^2) \det\left(I_{2m} - U \right),
\end{align}
where the $2m\times 2m$ matrix
\begin{equation}
    U \ldef D(G,\u) + D(G^*,\u^*)^\dag,
\label{eq:U}
\end{equation}
with the dual weight defined as $u_e^* \ldef (1 - u_e)/(1 + u_e)$. Here, the operator $D(G,\u)$, acting on the space of quadrilaterals, represents the weighted curl operators around the vertices in $V$, i.e., $D = \sum_{v \in V} D_v$, as depicted in Fig.~\ref{fig:square}. The matrix element
\begin{align}
    (D_v)_{q,q'}(u_e) = e^{i \gamma(q,q')/2 }  \frac{1 - u_e^2}{1 + u_e^2},
\label{eq:weight}
\end{align}
is nonzero when the quadrilaterals $q'$ and $q$ share a common edge $e$, with $q'$ positioned counterclockwise next to $q$. The phase $\gamma(q,q') = \angle{qvq'}$ denotes the angle wrapped around the vertex $v$. Similarly, the dual operator $D(G^*, \u^*)$ describes the curl around the vertices in $V^*$ (see red and blue cycles in Fig.~\ref{fig:square}).

Moreover, Eq.~\eqref{eq:U} suggests that the operator $U$ is self-dual under the dual transformation $\u\leftrightarrow\u^*$ and $G \leftrightarrow G^*$, leading to the manifest Kramers-Wannier (KW) duality in Eq.~\eqref{eq:mine1}, as $\zeta_F(G,\u) \sim \zeta_F(G^*, \u^*)$~\cite{cimasoni2012critical,cimasoni2015kac}. The operators $D_v$ and $D_{v^*}$ can be interpreted as local order and disorder operators, respectively, echoing the nonlocal disorder operator introduced by Kadanoff and Ceva~\cite{kadanoff1971determination}, which exchange under the dual transformation.

\mysec{Circle Theorem}
Here, we prove our main result, the circle theorem in Eq.~\eqref{eq:circle}, by showing that the operator $U$ is unitary, provided that all $u_e$ are real-valued. To do this, it is convenient to parameterize the weight as $u_e = \tan(\theta_e/2)$ for $\theta_e \in (-\pi,\pi)$. Then, the matrix weight in Eq.~\eqref{eq:weight} simplifies to
\(
\frac{1 - u_e^2}{1 + u_e^2} = \cos(\theta_e).
\)
Similarly, the dual weight is given by
\(
\frac{1 - {u_e^*}^2}{1 + {u_e^*}^2} = \frac{2u_e}{1+u_e^2} = \sin(\theta_e).
\)
It can be shown directly (see Appendix Section \ref{sec:circle}) that
\begin{equation}
   D(G,\u)  D (G,\u) ^\dag  = \mathrm{diag}\{\cos^2\theta_e\}.
\label{eq:Dcos}
\end{equation}
Similarly, applying the transformation $\u\to\u^*$, or equivalently $\theta_e \to \pi/2 - \theta_e$, we obtain
\begin{equation}
    D(G^*,\u^*)^\dag D (G^*,\u^*) = \mathrm{diag}\{\sin^2\theta_e\}.
\label{eq:Dsin}
\end{equation}
Now, consider the matrix $M \ldef D (G,\u) D (G^*,\u^*)$, whose elements are given by
\begin{equation}
    M_{q,q'} = M_{q',q} = e^{i \pi/2} \cos(\theta_e)\sin(\theta_e).
\notag
\end{equation}
Here, we use the fact that $\gamma(q,q'') + \gamma(q'',q') = \pi$, where $q''$ is the common quadrilateral between quadrilaterals $q$ and $q'$ in a counterclockwise ordering. This implies that $M$ is anti-Hermitian, i.e.,
\begin{equation}
     D (G,\u) D (G^*,\u^*) + (D (G,\u) D (G^*,\u^*))^\dag = 0.
\label{eq:crossing}
\end{equation}
Combining Eqs.~\eqref{eq:Dcos}--\eqref{eq:crossing}, we prove the unitarity of the operator $U$
\begin{equation}
    U U^\dag = I.
\notag
\end{equation}
Consequently, the eigenvalues of $U$, denoted as $\{z_1, z_2, \dots, z_{2m} \}$, satisfy $z_i = e^{i\phi_i}$, implying that they all lie on the unit circle, i.e., $|z_i|=1$. Substituting Eq.~\eqref{eq:mine1} into Eq.~\eqref{eq:path}, we obtain our main result:
\begin{equation}
    Z^2 = 2^{n} \prod_e\cosh(2\beta J_e) \prod_i (1-z_i),
\end{equation}
recovering the circle law in Eq.~\eqref{eq:circle}. Moreover, this new circle law holds for an arbitrary disorder realization $\u$. The quenched-averaged free energy, defined as $\beta \overline{f} \ldef  - \frac{1}{n}\overline{\ln Z}$, can be computed via the angular distribution of the zeros:
\begin{align}
   \beta \overline{f} = -\frac{1}{4} \left ( (2+d)  \ln 2 +d\, \overline{\ln\cosh(2\beta J)} \right. \notag \\
   \left.+ d \int_0^{2\pi} \ln(1-\cos(\phi)) g(\phi) d\phi \right),
\label{eq:freeenergy}
\end{align}
where $d \ldef 2m/n$ is the average degree of the graph, and $g(\phi) \ldef \frac{1}{2m} \overline{\sum_i \delta(\phi-\phi_i)}$ is the quenched-averaged angular distribution. As the zeros pinch the real axis at the critical point in the thermodynamic limit, it is important to understand the critical scaling of the leading zero $\phi_\mathrm{lead}$. If the leading zero vanishes as a power law,  
\begin{equation}
    \phi_\mathrm{lead}(T) \sim |T-T_c|^\Delta,
\label{eq:scaling}
\end{equation}
with exponent $\Delta$, then the free energy exhibits logarithmic singularities due to disorder. On the other hand, if the leading zero scales as a stretched exponential, the free energy would acquire non-trivial critical exponents, indicating a more complex critical behavior.

\begin{figure}[htp!]
  \includegraphics[width=1\linewidth]{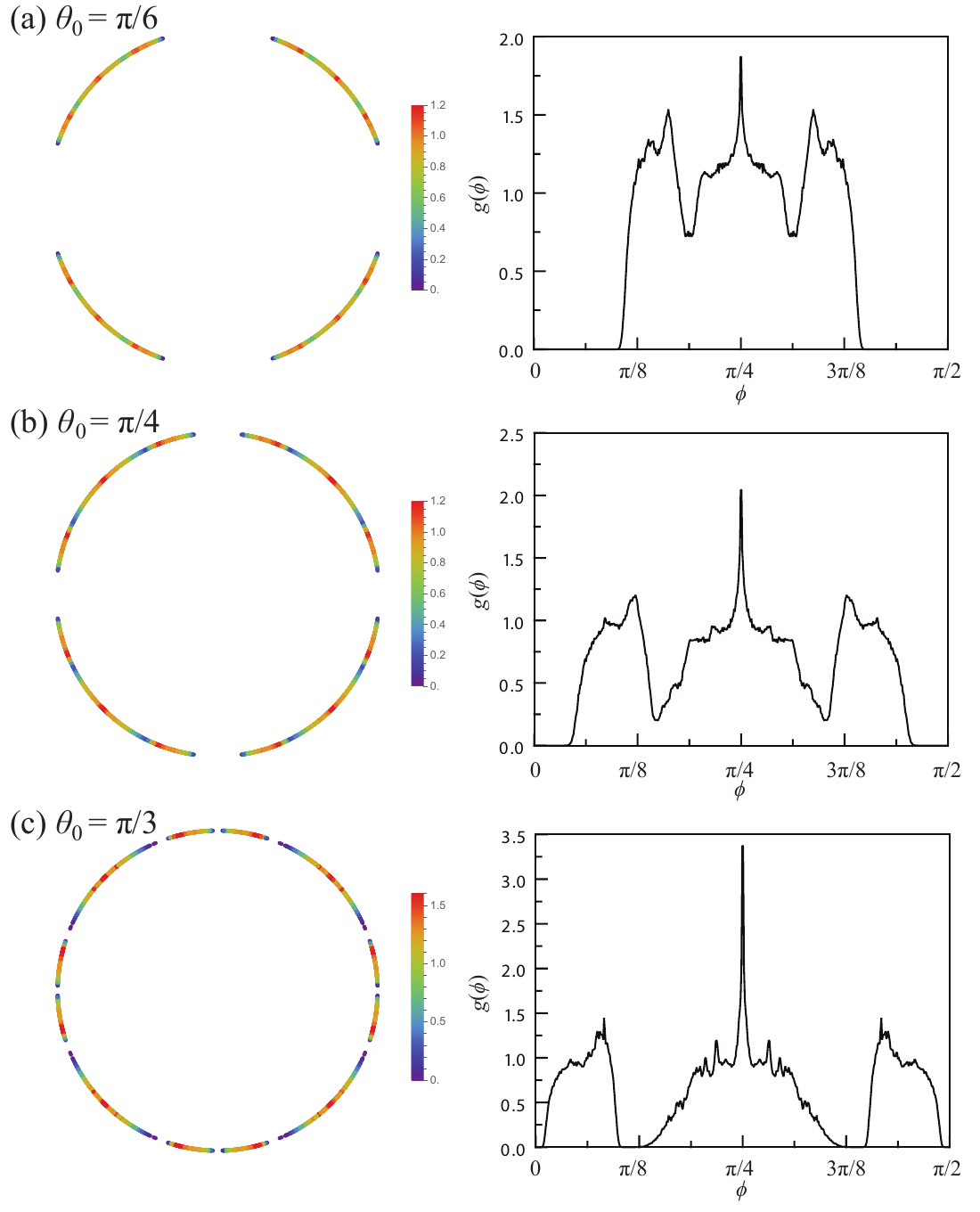}
  \caption{Circle law (left) and angular distributions (right) for the square lattice  $\pm J$ model with $p = 1/2$ at temperatures (a) $\theta_0 = \pi/6$ ($T\approx 1.6T_c^\mathrm{pure}$), (b) $\theta_0 = \pi/4 = \theta_c^\mathrm{pure}$ ($T=T_c^\mathrm{pure}$), and (c) $\theta_0 = \pi/3$ ($T\approx 0.67T_c^\mathrm{pure}$). 
}
  \label{fig:circle}
\end{figure}

\mysec{Implications}
We now turn to the implications of our circle law for Ising spin glasses. We focus on the square lattice; however, a similar analysis can be carried out for other planar graphs. Our discussion below primarily considers the $\pm J$ disorder on a square lattice, as $P(J) = (1-p)\delta(J+J_0) + p \delta(J-J_0)$, where $0 \leq p \leq 1$. Consequently, the bond disorder is parameterized as $u_{e} = u_0 \tau_{e}$, where $\tau_{e} = \pm 1$, and the curl operators simplify to  
\begin{align}
    D(G,\u) = e^{i\pi/4} \cos(\theta_0) d,\quad D(G^*,\u^*) = e^{i\pi/4} \sin(\theta_0)d^*, \notag
\end{align}
where both $d$ and $d^*$ are real matrices with elements $0$ and $\pm 1$, and $u_0 = \tan(\theta_0/2)$. Moreover, only $d^*$ accounts for the bond disorder $\tau_e$. Evidently, only the dual curl operator $D(G^*,\u^*)$ incorporates disorder, while the curl operator $D(G,\u)$ remains unaffected.

In the pure case ($p = 1$), where disorder is absent, it is straightforward to diagonalize $U$ via the Fourier transform, leading to the zeros  
\begin{align}
    \cos(2\phi) = \frac{1}{2}\sin(2\theta_0)(\cos(\varepsilon) + \cos(\eta)),
\label{eq:onsager}
\end{align}
where $(\varepsilon,\eta)$ represent the wave vectors of the square lattice. Together with Eq.~\eqref{eq:freeenergy}, we recover Onsager's free energy \cite{onsager1944crystal, kaufman1949crystal,kaufman1949crystalb}. Moreover, we find that the leading zero behaves as
\begin{align}
    \phi_\mathrm{lead}(T) = |\theta_0 - \theta_c^\mathrm{pure}| \sim |T-T_c|,\notag
\end{align}
where $\theta_c^\mathrm{pure} =  \pi/4$ corresponds to the critical temperature $T_c^\mathrm{pure}/J_0 =\tanh^{-1} [\tan(\theta_c^\mathrm{pure}/2)]$. This result agrees with our predictions that the zeros pinch the real axis precisely at the critical point, as suggested by Eq. \eqref{eq:scaling}

We now focus on the most interesting case, $p = 1/2$. Since $g(\phi)$ corresponds to the spectral density of a finite-dimensional random matrix $U$, analytical results can be difficult to obtain. Thus, we report numerical studies through a direct diagonalization of $U$ on a  $20\times20$ square lattice, averaged over $2,500$ disorder realizations. Figure~\ref{fig:circle} plots the zeros for three temperatures: above, at, and below $T_c^\mathrm{pure}$. We find that all zeros are located on the unit circle and symmetrically distributed across all four quadrants, validating the circle law for all cases, independent of temperature. Moreover, the zeros approach the real axis as the temperature decreases. As it is widely believed that the glass transition occurs at zero temperature~\cite{amoruso2003scalings,jorg2006strong}, we expect the zeros to pinch the real axis only at $T_c = 0$.

\begin{figure}[htp!]
  \includegraphics[width=1\linewidth]{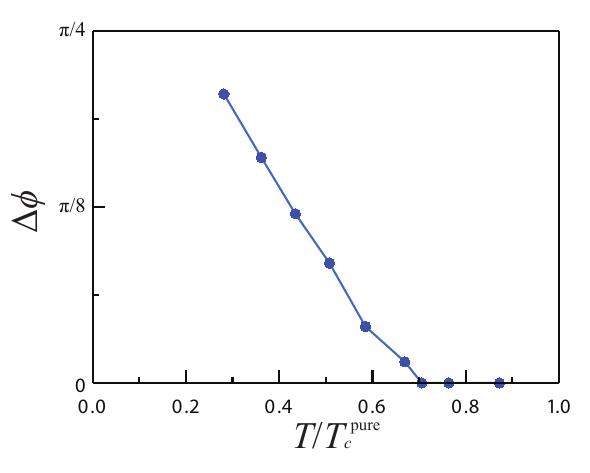}
  \caption{The spectral gap $\Delta\phi$ as a function of temperature $T$, suggesting a transition at $T_* \approx 0.7T_c^\mathrm{pure}$. 
}
  \label{fig:gap}
\end{figure}

Interestingly, we find that the angular distribution $g(\phi)$ in each quadrant exhibits two cusps for $T \geq T_c^\mathrm{pure}$, revealing a very different spectral density compared to the pure case in Eq. \eqref{eq:onsager}. More interestingly, for temperatures below a certain critical point $T_*$, the spectral density develops a gap, dividing the spectrum into a central band and two side bands.

Figure \ref{fig:gap} plots the gap size $\Delta\phi$ as a function of temperature, suggesting that $T_* \approx 0.7 T_c^\mathrm{pure}$. This suggests the presence of a crossover phase as a precursor before the system enters the true glassy state at zero temperature. Indeed, as the temperature decreases, all bands narrow, and the side bands move toward $0$ and $\pi/2$, leading to $\Delta \phi \to \pi/4$. Thus, the side bands eventually lead to the true glassy transition. This is consistent with the physical picture that their emergence is a precursor to the glassy transition.

In contrast, for Gaussian disorder, while we also find zeros moving toward the real axis as $T$ decreases, no such spectral gap has been observed (see Appendix Section~\ref{sec:gaussian}). This implies a fundamental distinction between the ground states of the $\pm J$ and Gaussian disorder models in 2D.

\mysec{Circle Theorem for Anyons}  
It is well known that the 2D Ising partition function is closely related to fermionic statistics. This is also evidenced by the negative sign for the odd winding number in Eq.~\eqref{eq:Fzeta}. Below, we show that the circle law can be generalized to fractional statistics of anyons. We start with a $q$-deformed zeta function
\begin{align}
\zeta_q(G,\u)^{-1} \equiv \prod_{[p]} \left( 1- q^{w(p)}\prod_{e\in p} u_e \right),
\label{eq:qZeta}
\end{align}
where $q \ldef e^{i 2 \pi s}$ lies on the unit circle, with $s$ being the spin. For half-integer spin ($q = -1$), we recover Eq.~\eqref{eq:Fzeta}, whereas for integer spin ($q=1$), which corresponds to the bosonic case, Eq.~\eqref{eq:qZeta} resembles the Ihara's graph zeta function, which has broad applications in number theory and graph theory \cite{ihara1966discrete,hashimoto1989zeta,sunada2006functions}.

Following a similar approach, we find (see Appendix Section~\ref{sec:deform})
\begin{equation}
    \zeta_q(G,\u)^{-1} =\frac{\prod_{e}(1-qu_e^2)}{ (1-q)^{n}}\det\left(I - U_q\right),
\notag
\end{equation}
where $U_q \ldef D_q + {D_q^*}^\dag$, with matrix elements given by  
\begin{align}
    (D_q)_{q,q'} = e^{i\gamma(q,q')s}\frac{1-u_e^2}{1-qu_e^2},\;
    (D_q^*)_{q,q'} = e^{i\gamma^*(q,q')s}\frac{(1-q)u_e}{1-qu_e^2}.
    \notag
\end{align}
In the limit \(q\to 1\), this formula recovers the well-known {\it Bass formula}~\cite{bass1992ihara,foata1999combinatorial,terras2010stroll} for the Ihara zeta function (see Appendix Section~\ref{sec:ihara}).

Interestingly, the matrix $U_q$ is also unitary for $u_e \in \mathbf{R}$. The proof follows a similar structure to the one outlined above, relying on the following two key identities:  
\begin{equation}
    \left | \frac{1-u_e^2}{1-q u_e^2} \right|^2 + \left | \frac{(1-q)u_e}{1-q u_e^2} \right|^2 = 1,
\notag
\end{equation}
which generalizes Eqs.~\eqref{eq:Dcos}--\eqref{eq:Dsin}, and the term
\begin{equation}
    q^{1/2}\frac{1-u^2}{1-q u_e^2} \left( \frac{(1-q)u_e}{1-q u_e^2} \right)^*  
\notag
\end{equation}
is purely imaginary, generalizing Eq.~\eqref{eq:crossing}. Consequently, the circle theorem remains valid for arbitrary \( |q| = 1 \), extending its applicability to anyonic statistics.

\mysec{Dicussion}
To summarize, we have introduced a new circle theorem for Ising spin glasses on planar graphs under an arbitrary quenched disorder. This provides a new tool to analyze spin-glass phase transitions within the Lee-Yang framework. Moreover, we extend this theorem to anyonic statistics through a $q$-deformation. However, the precise physical realization of this deformation remains an open question. Unlike the standard Ising case, where the connection to fermionic statistics is well understood, extending the theorem to arbitrary \( q = e^{i2\pi s} \) suggests a potential connection to a broader class of models exhibiting fractional statistics~\cite{fradkin1980disorder}, though this connection remains speculative and requires further investigation.

Another open question concerns the limitation of our approach to the zero external field case ($z=1$). Unlike Lee-Yang theory, which applies to general $z = e^{-2\beta h}$, our result lacks this variable dependence. This raises a natural question: for what values of $z$ do the partition function zeros vanish? Equivalently, what does the product $\prod_i (z - z_i)$ correspond to for an arbitrary value of $z$? Clarifying this point would deepen our understanding of the discovered circle law and may also provide potential avenues for experimental validation.  

When applied to 2D Ising spin glasses, we find that the leading zero approaches the real axis as the temperature decreases, indicating a zero-temperature spin-glass phase. Furthermore, the emergence of a finite-temperature spectral gap for the $\pm J$ disorder, which is absent in the Gaussian disorder case, suggests that the former case introduces additional degeneracies or domain-wall structures that do not appear in the latter systems. This hints at a fundamental distinction between the ground-state properties of these two disorder types.

Although the true spin-glass transition in two dimensions is believed to occur strictly at zero temperature, the onset of a spectral gap at a finite temperature may represent a nontrivial precursor phase associated with the formation of metastable low-energy states. This behavior is reminiscent of the Griffiths singularities~\cite{griffiths1969nonanalytic}, where rare locally ordered regions emerge above the critical temperature. Whether this spectral gap phenomenon and Griffiths singularities are fundamentally related remains an open question.

Moreover, it is worth noting that spectral cusps and gaps similar to those observed in our numerical results also appear in certain correlated random matrix ensembles \cite{alt2020dyson}. This suggests that random matrix theory may provide further insights into the underlying nature of these transitions, a direction we leave for future exploration. As the primary focus of this paper is to establish the new circle theorem, our numerical investigation remains preliminary and supportive, with many open questions for future work. In particular, a more detailed study of critical scaling, finite-size effects, and universality classes would be valuable extensions.

Nevertheless, our theorem has potential implications beyond condensed matter physics. Spin glass models play a crucial role in optimization problems, machine learning, and quantum computing, and the geometric characterization of partition function zeros in such contexts could offer new theoretical perspectives. Additionally, the connection between our zeta function formulation and mathematical physics is noteworthy, as our theorem generalizes the well-known graph zeta function. In this sense, our circle law serves as an analog to the Riemann Hypothesis (RH), partially addressing longstanding questions about potential connections between the Lee-Yang theory and RH.

\appendix

\section{Derivation of the Self-dual Formula}\label{sec:self}
In this section, we provide a concise review of the self-dual formula discovered in Ref.~\cite{song2024kramers}. We consider the Ising spin glass defined on a planar graph $G = (V, E)$ with $n = |V|$ and $m = |E|$, where the weights $\u = \{ u_e \mid e\in E\}$ are assigned to each edge $e \in E$, with $u_e \ldef \tanh(\beta J_e)$. The Kac-Ward formula is given by 
\begin{equation}\label{eqS:Fedge}
\zeta_F(G,\u)^{-1} = \det \left(I_{2m} - T_{KW}(\u)\right),
\end{equation}
where the Kac-Ward matrix is defined over the set of oriented edges, $E \cup \bar{E}$, effectively doubling the number of edges. Given an oriented edge $e \in E$, we denote its inverse as $\bar{e} \in \bar{E}$. Consequently, we extend the weight assignment to oriented edges by setting $u_{\bar{e}} = u_e$.  

\begin{figure}[thb!]
  \includegraphics[width=1\linewidth]{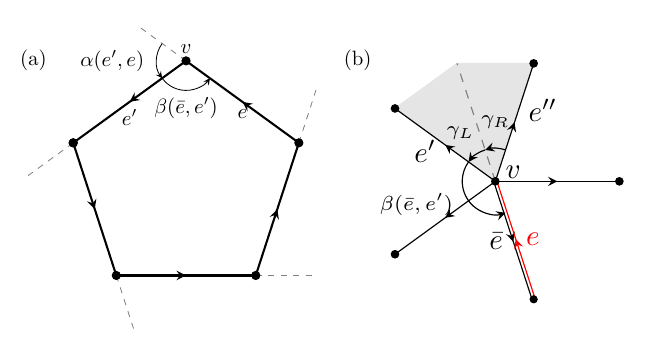}
  \caption{(a) The exterior angle $\alpha(e',e)$ and the interior angle $\beta(\bar{e}, e')$ for the KW operator satisfy $\alpha(e',e) = \pi - \beta(\bar{e}, e')$. (b) The angles between two neighboring edges attached to a quadrilateral satisfy $\beta(\bar{e},e') = \beta(\bar{e},e'') - \gamma_L(e') - \gamma_R(e'')$. Figure adapted from Ref.~\cite{song2024kramers}. }
  \label{fig:angle}
\end{figure}

Given the setup above, the Kac-Ward matrix is given by
\begin{align}
    (T_{KW}(\u))_{e',e} = e^{i\alpha(e',e)/2}  u_{e} = i e^{-i \beta(\bar{e},e')/2} u_{e}, 
\end{align}
if edge $e'$ follows $e$ without backtracking, where $\alpha(e',e)$ is the exterior angle from $e$ to $e'$, and $\beta(\bar{e}, e')$ is the interior angle between $e'$ and its inverse $\bar{e}$ (Fig.~\ref{fig:angle}a). This condition ensures that the sum of half-exterior angles contributes to a total phase change of $\pi$ over a cycle, effectively capturing the fermionic sign.

\begin{figure}[!thb]
  \includegraphics[width=1\linewidth]{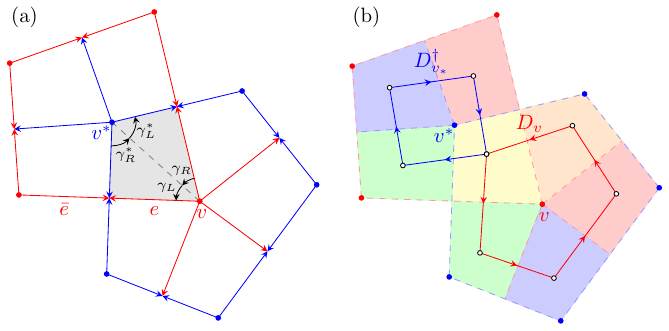}
\caption{(a) The embedding of both $G$ and its dual $G^*$. The quadrilateral $q$ is delineated by a vertex $v$ and a neighboring dual vertex $v^*$, along with their respective edges. The relationships $\gamma_L + \gamma_R^* = \gamma_R + \gamma_L^* = \pi/2$ are satisfied. (b) The local order and disorder operators $d_v$ and $d_{v^*}^\dag$ for quadrilaterals. Each operator acts as a curl operator around the vertex $v$ and the dual $v^{*}$, respectively. Figure adapted from Ref.~\cite{song2024kramers}.}
\label{fig:embed}
\end{figure}

To derive our main result, we embed the dual graph $G^*$ over $G$. In this arrangement, each vertex of $G$ is located inside a face of $G^*$, and vice versa; each edge in $G$ intersects its corresponding edge in $G^*$. For technical convenience, we require these intersections to be perpendicular. Note that the embedding need not be isoradial in general. Figure~\ref{fig:embed}a illustrates such an embedding, where red and blue colors represent $G$ and $G^*$, respectively. 

A crucial element in our construction involves quadrilaterals (gray domain in Fig.~\ref{fig:embed}a). Each quadrilateral is formed by two neighboring edges in both $G$ and $G^*$, along with a vertex pair $(v,v^*)$ where $v \in V$ and $v^* \in V^*$. We denote the angles associated with the left and right edges of $v$ and $v^*$ as $\gamma_L$ and $\gamma_R$, and $\gamma_L^*$ and $\gamma_R^*$, respectively. These angles satisfy the relations $\gamma_L + \gamma_R^* = \gamma_R + \gamma_L^* = \pi/2$ as depicted in Fig.~\ref{fig:embed}a. The collection of $2m$ quadrilaterals, which collectively tile the entire plane, plays a critical role in our new formulation.

We introduce the gauged successor operator 
\begin{equation}
   S(\u) = T_{KW}(\u) - i J(\u),
\end{equation}
by appending an additional element between edge $e$ and its inverse $\bar{e}$. Here, $J_{e,\bar{e}} = u_e$ is nonzero only for elements between the edge and its inverse. The additional phase factor is given by $ie^{-i\beta(\bar{e}, e)} = -i$.

Following the strategy outlined in Refs.~\cite{cimasoni2012critical,cimasoni2015kac}, we introduce the operator $Q$ between neighboring edges $e$ and $e'$ that share a common starting vertex $v$. Specifically, for each quadrilateral $q$, the operator $Q$ maps its right edge $e$ to the left edge $e'$, while inducing a phase shift, 
\begin{equation}
    Q_{e,e'} = e^{i(\gamma_{L}+\gamma_{R})/2}.
\end{equation}
As illustrated in Fig.~\ref{fig:angle}b,  $Q_{e',e''} e^{-i\beta(e'',e)/2}  = e^{-i\beta(e',e)/2}$ if $e' \neq \bar{e}$. However, the fermionic nature exhibits a nontrivial monodromy, resulting in a branch cut after a $2\pi$ rotation. This observation leads to the discontinuity
\begin{align}\label{eqS:R}
S(\u) - QS(\u) = -2iJ(\u).
\end{align}

Since the operator $Q$ acts on edges attached to quadrilaterals, it naturally factors as $$Q = L^t R,$$ where $L$ and $R$ are $2m\times 2m$ matrices associating each quadrilateral $q$ with its left and right edges, respectively, with the matrix elements 
\begin{equation}\label{eqS:LR}
L_{q,e} = e^{i\gamma_L(q)/2},\; R_{q,e} = e^{i\gamma_R(q)/2}.    
\end{equation}
Building on this factorization, we find
\begin{align}
&\det(I-Q)\det(I-T_{KW}(\u)) = \det\left(I+iJ(\u) -  Q(I-iJ(\u))\right) \notag\\
& = \det(I+iJ(\u))\det\left(I - R  (I-iJ(\u)) (I+iJ(\u))^{-1} L^t \right) \notag\\
& =  \prod_e(1+u_e^2) \det\left( I -  R K(\u) L^t -   Re^{-i\pi/2}K^*(\u)J L^t \right), \notag
\end{align}
where $K(\u)$ and $K^*(\u)$ are diagonal matrices with 
\begin{equation}
    K_{ee} = K_{\bar{e}\bar{e}} = \frac{1-u_e^2}{1+u_e^2}, \; K^*_{ee} = K^*_{\bar{e}\bar{e}} =  \frac{2u_e}{1+u_e^2}. \notag
\end{equation}
It is worth noting that $K^*(\u) = K(\u^*)$ under the dual transformation $u_e \to u_e^* \ldef (1-u_e)/(1+u_e) $. Finally, we obtain
\begin{align}\label{eqS:mine1}
    \zeta_F^{-1}(G,\u) = 2^{-n} \prod_{e\in E}(1+u_e^2) \det\left(I_{2m} - D(G,\u)- D(G^*, \u^*)^\dag \right).
\end{align}

\section{$q$-Deformation}\label{sec:deform}
In this section, we follow a similar approach as in Section~\ref{sec:self}, applying it to the $q$-deformed zeta function
\begin{align}\label{eqS:qzeta}
\zeta_F(G,\u)^{-1} \equiv \prod_{[p]} \left( 1- q^{w(p)}\prod_{e\in p} u_e \right),
\end{align}
where $q \ldef \exp(i2\pi s)$, with $s$ representing the spin. In the case $s = 1/2$, i.e., $q = -1$, we recover the standard definition of the fermionic zeta function $\zeta_F$. Similarly, it can be computed via a determinant as
\begin{equation}\label{eqS:qedge}
\zeta_q(G,\u)^{-1} = \det \left(I_{2m} - T_q(\u)\right),
\end{equation}
where the edge transfer matrix $T_q$ is defined as 
\begin{align}
    (T_q(\u))_{e',e} = e^{i\alpha(e',e)s}  u_{e} = i e^{-i \beta(\bar{e},e')s} u_{e}.
\notag
\end{align}
Using the same approach as in Section~\ref{sec:self}, we introduce the successor matrix 
\begin{equation}
   S_q(\u) = T_q(\u) + q^{-1/2} J(\u),
\notag
\end{equation}
along with the $Q_q$ operator, defined as 
\begin{equation}
    (Q_q)_{e,e'} = e^{i(\gamma_{L}+\gamma_{R})s}.
\notag
\end{equation}
This operator factorizes as 
\begin{equation}
    Q_q = L_q^t R_q,
\notag
\end{equation}
with $(L_q)_{q,e} = e^{i \gamma_L(q) s}$ and $(R_q)_{q,e} = e^{i \gamma_R(q) s}$. 
This generalizes Eq.~\eqref{eqS:R} as 
\begin{align}\label{eqS:Rq}
S_q(\u) - Q_qS_q(\u) =  (q^{-1/2} - q^{1/2})J(\u).
\end{align}
Similarly, we find
\begin{align}
&\det(I-Q_q)\det(I-T_q(\u)) \notag\\
&= \det\left(\left(I+ q^{1/2} J(\u)\right) - Q_q\left(I+ q^{-1/2}J(\u)\right) \right) \notag\\
& =\det\left(I+q^{1/2}J(\u)\right)  \notag\\
&\times \det\left(I - R_q \left(I+ q^{-1/2}J(\u)\right)\left(I+ q^{1/2}  J(\u) \right)^{-1}  L_q^t \right).
\notag
\end{align}
We use the identities $\det(I-Q_q) = (1-q)^n$, $\det\left(I+q^{1/2}J(\u)\right) = \prod_e (1-qu_e^2)$, 
and
\begin{equation}
\left(I+ q^{-1/2}J(\u)\right)\left(I+ q^{1/2}  J(\u) \right)^{-1}   = K_q(\u) + K_q^*(\u) J,
\notag
\end{equation}
where $K_q$ and $K_q^*$ are diagonal matrices, with
\begin{equation}
    (K_q)_{ee} = (K_q)_{\bar{e}\bar{e}} =\frac{1-u_e^2}{1-qu_e^2}, \; (K_q^*)_{ee} = (K_q^*)_{\bar{e}\bar{e}} =  \frac{(1-q)u_e}{1-qu_e^2}. \notag
\end{equation}
Combining these results, we obtain the $q$-analogue of Eq.~\eqref{eqS:mine1}:
\begin{equation}
    \zeta_q(G,\u)^{-1} =\frac{\prod_{e}(1-qu_e^2)}{ (1-q)^{n}}\det\left(I - U_q\right),
\label{eqS:mineq}
\end{equation}
where $U_q(\u) = D_q(\u) + D_q^*(\u)^\dag$, with $D_q(\u) \ldef R_qK(\u)L_q^{t}$ and $D_q^*(\u) \ldef R_q q^{-1/2} K^*(\u)J L_q^t$, having matrix elements
\begin{align}
    (D_q)_{q,q'} = e^{i\gamma(q,q')s}\frac{1-u_e^2}{1-qu_e^2},\;
    (D_q^*)_{q,q'} = e^{i\gamma^*(q,q')s}\frac{(1-q)u_e}{1-qu_e^2}.
    \notag
\end{align}

\section{Connection to Ihara Zeta Function}\label{sec:ihara}
In this section, we consider the bosonic case in the limit $q\to 1$. In this case, Eq.~\eqref{eqS:qzeta} reduces to the Ihara zeta function \cite{ihara1966discrete,sunada2006functions},
\begin{align}\label{eqS:iharazeta}
\zeta_1(G,\u)^{-1} \equiv \prod_{[p]} \left( 1- \prod_{e\in p} u_e \right).
\end{align}
The determinant formula in Eq.~\eqref{eqS:qedge} simplifies to
\begin{equation}\label{eqS:hedge}
\zeta_1(G,\u)^{-1} = \det \left(I_{2m} - T_1(\u)\right),
\end{equation}
where $T_1$ is Hashimoto's edge adjacency operator \cite{hashimoto1989zeta}. 

Next, we take the limit $q\to 1$ in Eq.~\eqref{eqS:mineq}. Setting $\epsilon \ldef 1-q$, we then have 
\begin{equation}
    D_q(\u) = D_1 - \epsilon D_1(\u)' + O(\epsilon^2),\; D_q^*(\u) = -\epsilon D_1^*(\u)' + O(\epsilon^2),
\notag\end{equation}
where $(D_1)_{q,q'} = 1$ and  
\begin{equation}
    D_1(\u)'_{q,q'} = \frac{u_e^2}{1-u_e^2} + \frac{\gamma(q,q')}{2\pi},\; D_1^*(\u)'_{q,q'}= -\frac{u_e}{1- u_e^2},
\notag\end{equation}
if $q$ and $q'$ are consecutive.
Let $H_0 \ldef I_{2m} - D_1$ and $H_1 = D_1(\u)' + D_1^*(\u)'$. Then, Eq.~\eqref{eqS:mineq} becomes
\begin{equation}
    \zeta_1(G,\u)^{-1}  = \prod_e(1-u_e^2) \lim_{\epsilon\to 0} \epsilon^{-n} \det M(\u),
\notag\end{equation}
where $M(\u) \ldef I - U_q(\u) = H_0 + \epsilon H_1(\u) + O(\epsilon^2)$.

Since $H_0$ corresponds to the unweighted curl operator, it is block diagonalized into $n$ blocks, each associated with a vertex $v$, and each block contributes a zero mode given by
\begin{equation}
    |v \rangle\ldef \frac{1}{\sqrt{d_v}} \sum_{v' \in N(v)} e_{v'},
\notag\end{equation}
which forms an $n$-dimensional null space $V_0$. Here, $d_v$ denotes the degree of vertex $v$, $N(v)$ is the set of neighbors of $v$, and $e_{v}$ is the unit vector with one at $v$ and zero elsewhere. Denoting the complementary space of $V_0$ as $V_\perp$, the $2m$-dimensional Hilbert space decomposes as $V = V_0 \oplus V_\perp$. Using this representation, we obtain
\begin{equation}
\det M = \det
    \begin{bmatrix}
M_{00} & M_{0\perp} \\
M_{\perp 0} & M_{\perp\perp}
    \end{bmatrix} = \det(M_{00})\det(M_{\perp\perp}-M_{\perp 0}M_{00}^{-1}M_{0\perp}),
\notag\end{equation}
where $M_{00} = \epsilon H_1^{00} + O(\epsilon^2)$, and $M_{0\perp}$ and $M_{\perp 0}$ are leading-order terms in $\epsilon$. Since $M_{\perp 0}M_{00}^{-1}M_{0\perp}$ is also leading order in $\epsilon$, it follows that $M_{\perp\perp}-M_{\perp 0}M_{00}^{-1}M_{0\perp} = H_0^{\perp\perp} + O(\epsilon)$. Taking the limit, we obtain
\begin{equation}
    \lim_{\epsilon\to 0} \epsilon^{-n} \det M = \det(H_0^{\perp\perp})\det(H_1^{00}),
\notag\end{equation}
where $H_0^{\perp\perp}$ is the projection of $H_0$ onto the complementary space. Its determinant can be computed as
\begin{equation}
   \det(H_0^{\perp\perp}) = \prod_v \prod_{k=1}^{d_v-1} \left(1- e^{i 2\pi k/d_v}\right)  = \prod_v d_v.
\notag\end{equation}
Now, we consider the operator $H_1^{00}$, which is $H_1$ projected onto the null space. We find that 
\begin{align}
    \langle v| H_1 | v' \rangle = &\frac{1}{d_v}\left(\sum_{v'' \in N(v)}\frac{u_{(vv'')}^2}{1-u_{(vv'')}^2} + 1 \right) \delta_{vv'}  \notag\\
    &- \frac{1}{\sqrt{d_v d_{v'}}} \frac{u_{(vv')}}{1- u_{(vv')}^2}  A_{v,v'},
\notag\end{align}
where $A$ is the adjacency matrix. Therefore,
\begin{equation}
    \det(H_0^{\perp\perp}) = \prod_{v} (d_v)^{-1} \det (I_{n}-\mathbf{A}(\u)+\mathbf{Q}(\u)),
\notag\end{equation}
where $\mathbf{Q}(\u)$ is a diagonal matrix and $\mathbf{A}(\u)$ is the weighted adjacency matrix. Their matrix elements are given by
\begin{equation}
    \mathbf{Q}_{v,v} = \sum_{v' \in N(v)}\frac{u_{(vv')}^2}{1-u_{(vv')}^2},\;  \mathbf{A}_{v,v'} = \frac{u_{(vv')}}{1- u_{(vv')}^2}A_{v,v'}.
\notag\end{equation}
Combining everything, we obtain
\begin{equation}
    \zeta_1(G,\u)^{-1} = \prod_e(1-u_e^2)\det (I_{n}-\mathbf{A}(\u)+\mathbf{Q}(\u)).
\end{equation}
In the case of uniform weight $u_e = u$, the equation above reduces to the well-known Bass formula~\cite{bass1992ihara,foata1999combinatorial,terras2010stroll},
\begin{equation}
      \zeta_1(G,u)^{-1} = (1-u^2)^{m-n}\det (I_{n}-u A+ u^2 Q),
\end{equation}
where $Q$ is a diagonal matrix with $Q_{vv} = d_v-1$.

\section{Circle Theorem}\label{sec:circle}
\begin{figure}[thb!]
\centering
  \includegraphics[width=0.5\linewidth]{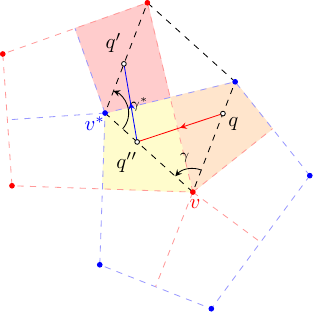}
  \caption{Illustration of \( D_q D_q^* \), where \( D_q \) and \( D_q^* \) are represented by red and blue arrows, respectively. As the sum of two adjacent interior angles of a parallelogram (formed by black dashed lines) equals \( \pi \), the phases satisfy \( \gamma(q,q'')+\gamma^*(q'',q') = \pi \).}
  \label{fig:cross}
\end{figure}
In this section, we prove the circle theorem Eq.~\eqref{eqS:mineq} by showing that the operator \( U_q \) is unitary, provided that all \( u_e \) are real-valued. 
\begin{equation}
 U_q U_q^\dag = ( {D_q} +{D_q^*}^\dag  ) ( {D_q}^\dag +{D_q^*} ) 
\end{equation}
It can be directly shown that
\begin{equation}
      D_q {D_q}^\dag  = K_q K_q^\dag,\; {D_q^*}^\dag  D_q^* = K_q^* {K_q^*}^\dag.
\notag
\end{equation}
Using the fact that \( 2-|1-q|^2=q+q^* \), we obtain
\begin{equation}
    \left( 1-u_e^2 \right)^2 + \left|1-q\right|^2 u_e^2 = \left|1-q u_e^2\right|^2.
\end{equation}
Consequently, 
\begin{equation}
    D_q {D_q}^\dag +  {D_q^*}^\dag  D_q^*= I_{2m}.
\end{equation}

Now, consider the cross term 
\begin{equation}
    (D_q D_q^*)_{q,q'} = e^{i \pi s}   \frac{1-u_e^2}{1-q u_e^2} \left( \frac{(1-q)u_e}{1-q u_e^2} \right)^* .
\notag
\end{equation}
Here, we use the fact that \( \gamma(q,q'') + \gamma(q'',q') = \pi \), where \( q'' \) is the common quadrilateral between quadrilaterals \( q \) and \( q' \) in a counterclockwise ordering (see Fig.~\ref{fig:cross}).  

Since 
\begin{equation}
 q^{1/2}  \frac{1-u_e^2}{1-q u_e^2} \left( \frac{(1-q)u_e}{1-q u_e^2} \right)^* = i 2\sin(\pi s)\frac{(1-u_e^2)u_e}{\left|1-q u_e^2\right|^2} 
\end{equation}
is purely imaginary, with \( q^{1/2} = e^{i\pi s} \), we find 
\begin{equation}
    D_q D_q^* +  (D_qD_q^*)^\dag = 0.
\label{eqS:crossing}
\end{equation}
Combining these results, we prove the unitarity of the operator \( U_q \):
\begin{equation}
     U_q U_q^\dag  = I_{2m}.
\end{equation}

\section{Gaussian Disorder Model}\label{sec:gaussian}
\begin{figure}[thb!]
  \includegraphics[width=1\linewidth]{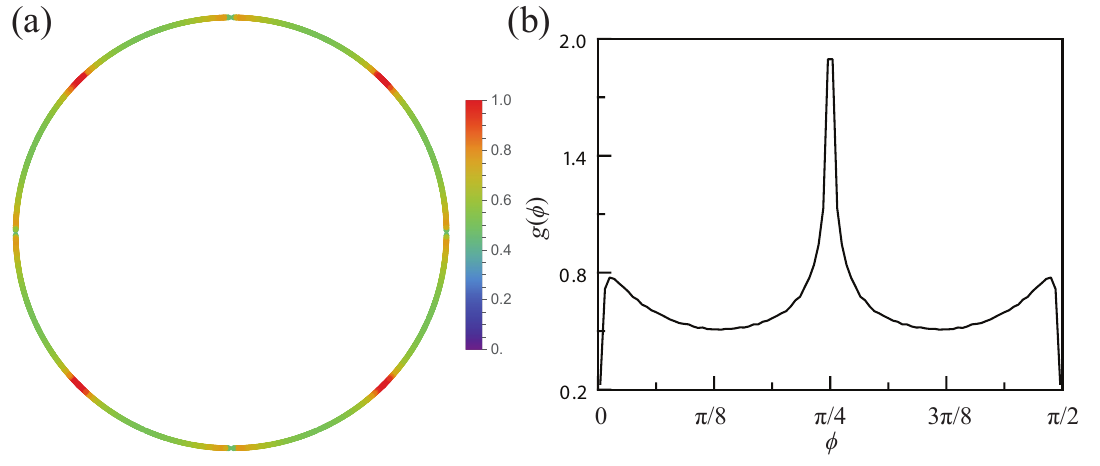}
  \caption{The square lattice Gaussian disorder model at temperature \( T = \sigma_J \), showing (a) the zero distribution and (b) the angular distribution.}
  \label{fig:EA}
\end{figure}

In this section, we analyze the angular distribution of zeros for the Gaussian disorder model in the square lattice Ising spin glass, given by  
\begin{equation}
    P(J) = \frac{1}{\sqrt{2\pi}\sigma_J} e^{-\frac{J^2}{2\sigma_J^2}},
\end{equation}
where \( \sigma_J \) controls the variance of the disorder. Figure~\ref{fig:EA} presents the zero distribution and angular distribution at the temperature \( T = \sigma_J \). We find that while the circle law holds, no cusps or band gaps are observed, indicating a fundamental difference from the $\pm J$ disorder model. The zeros are very close to the real axis, yet they do not touch it.

\end{document}